\documentclass[prx,twocolumn,longbibliography,superscriptaddress,preprintnumbers]{revtex4-2}
\usepackage[colorlinks,bookmarks=true,citecolor=blue,linkcolor=blue,urlcolor=blue]{hyperref}
\usepackage{dcolumn,graphicx,amsfonts,amsthm,bm,color,appendix,float,tabularx}
\usepackage{braket}
\usepackage[normalem]{ulem}
\usepackage[version=4]{mhchem}
\usepackage{siunitx}
\usepackage{amsmath}
\usepackage{amssymb}
\usepackage{CJKutf8}
\usepackage{multirow}
\usepackage{booktabs}

\usepackage[dvipsnames]{xcolor}
\definecolor{pal0}{rgb}{0.8941, 0.102 , 0.1098}
\definecolor{pal1}{rgb}{0.2157, 0.4941, 0.7216}
\definecolor{pal2}{rgb}{0.302 , 0.6863, 0.2902}
\definecolor{pal3}{rgb}{0.5961, 0.3059, 0.6392}
\definecolor{pal4}{rgb}{1.    , 0.498 , 0.    }

\newcommand{\dn}[1]{\left|\left| #1 \right|\right|}

\renewcommand{\v}[1]{\boldsymbol{#1}}

\newcommand{\bz}{\mathrm{BZ}}

\DeclareMathOperator{\sgn}{sgn}

\newcommand{\PRLsec}[1]{\emph{#1---}}

\usepackage{enumitem,amssymb}
\newlist{todolist}{itemize}{2}
\setlist[todolist]{label=$\square$}
\usepackage{pifont}

\newcommand{\tcell}[1]{\parbox[c]{0.7\linewidth}{\raggedright\vspace{3pt} #1 \vspace{3pt}}}

\usepackage{blindtext}

\begin{document}

\title{
Crystals Caught Doping: Metallic Wigner Crystals in Rhombohedral Graphene
}

\author{Junkai Dong (\begin{CJK*}{UTF8}{bsmi}董焌\end{CJK*}\begin{CJK*}{UTF8}{gbsn}锴\end{CJK*})}
\thanks{These authors contributed equally.}
\affiliation{Department of Physics, Harvard University, Cambridge, MA 02138, USA}

\author{Tomohiro Soejima (\begin{CJK*}{UTF8}{bsmi}副島智大\end{CJK*})}
\thanks{These authors contributed equally.}
\affiliation{Center for Computational Quantum Physics, Flatiron Institute,
162 5th Avenue, New York, NY 10010, USA}
\affiliation{Center for Quantum Phenomena, Department of Physics,
New York University, 726 Broadway, New York, New York 10003, USA}

\author{Daniel E. Parker}
\affiliation{Department of Physics, University of California at San Diego, La Jolla, California 92093, USA}

\author{Ashvin Vishwanath}
\affiliation{Department of Physics, Harvard University, Cambridge, MA 02138, USA}

\begin{abstract}
Nearly a century after Wigner's initial proposal, electron crystals are now a topic of intense experimental and theoretical interest. However, most proposed crystalline phases are commensurate and therefore become insulating in the presence of even weak pinning.
In this work we discuss when a 
commensurate Wigner crystal will spontaneously self dope and develop itinerant carriers, giving rise to an incommensurate and thus metallic Wigner crystal (MWC). We develop a general criterion for the instability of the commensurate crystal which involves the competition between the charge gap at commensurability and a ``packing bias'' whose sign selects whether electron or hole doping is preferred.
We then apply these insights to rhombohedral multilayer graphene, where calculations for commensurate crystals reveal instabilities towards self-doping.
Carrying out self-consistent Hartree-Fock over the landscape of incommensurate crystals reveals the phase diagram, where a broad MWC phase appears directly adjacent to an insulating Wigner crystal phase.
Recent observations of an island of reversed Hall conductance near a putative Wigner crystal phase in rhombohedral graphene are naturally explained by our theory.
\end{abstract}

\maketitle

\PRLsec{Introduction} 
Wigner crystals are a preeminent example of the competition between interaction driven localization and kinetic energy driven itineracy~\cite{wigner_interaction_1934}. Historically, this problem has been studied in the jellium model, where electrons with a quadratic dispersion interact via Coulomb forces in a neutralizing background~\cite{Giuliani_Vignale_2005}.
At large interaction strengths, the electrons spontaneously break continuous translation symmetry to form a lattice commensurate with the density, where each electron occupies its own lattice site~\cite{tanatar1989ground,Drummond_Phase_Diagram,lozovik_crystallization_1975, grimes_evidence_1979, andrei_observation_1988, yoon_wigner_1999, ma_thermal_2020, smolenski_signatures_2021,zhou_bilayer_2021, yang_experimental_2021,falson_competing_2022,sung_observation_2023,tsui_direct_2023,xiang_quantum_2024, zhou2025electroniccrystalslayeredmaterials}. 

Experiments on rhombohedral multilayer graphene (RMG)~\cite{zhou2021half,hallcrystal2022,lu2024fractional, waters2025chern,xiang2025continuouslytunableanomaloushall} have set off much recent activity on electron crystals.
Here, bands can be made extremely narrow so that interaction effects dominate, and at low carrier density electrons can crystallize on length scales far exceeding the atomic lattice spacing.
However, the relevant electronic states of RMG differ in two important ways from the jellium model. 
First, their non-interacting dispersion develops a ``Mexican hat" profile whose central peak increases with displacement field~\cite{zhang2010band}. Second, their multilayer spinor structure carries non-trivial quantum geometry.
Such differences can substantially modify the physics of electron crystallization, exemplified by the case of $\lambda$-jellium~\cite{AHC3, Tan_parent_berry}, where a recent variational Monte Carlo study found that adding non-trivial quantum geometry to jellium reduces the critical density for crystallization by up to an order of magnitude~\cite{valenti2025quantumgeometrydrivencrystallization}.
Moreover, quantum geometry also gives rise to entirely new crystalline phases, including ``halo'' Wigner crystals~\cite{joy2025chiral, AHC3} and anomalous Hall crystals (AHCs), which combine topology with crystallization~\cite{AHC_Yahui,AHC1,kivelson_cooperative_1986,halperin_compatibility_1986,kivelson_cooperative_1987,tesanovic_hall_1989, AHC2, Zhihuan_Stability, zhou2024newclassesquantumanomalous, guo2025correlationstabilizedanomaloushall, crepel2025efficient, bernevig2025berrytrashcanmodelinteracting,Zeng_sublattice_structure, Tan_parent_berry, Tan_FAHC, Zeng_sliding,  AHC3, AHC4, desrochers2025elasticresponseinstabilitiesanomalous, patri2025family, valenti2025quantumgeometrydrivencrystallization, lu2026generalmanybodyperturbationframework,desrochers2026electronic, tan2025ideallimitrhombohedralgraphene}. 
However, these states are all commensurate, with an integer number of electrons per unit cell.

A natural next question is: must Wigner crystals always be commensurate? The alternative is an incommensurate Wigner crystal, where continuous translations are spontaneously broken, but the number of particles per crystalline unit cell is not an integer. Even with a pinning potential such crystals are metallic, effectively realizing a translation-symmetry-broken metal. We refer to such states as \textit{metallic Wigner crystals} (MWC). 
An early incarnation of this idea is the spin-density wave instability of the Fermi liquid state proposed by Overhauser~\cite{overhauser1960giant, overhauser1962spin},  which is borne out in the incommensurate density wave states found by mean field calculations in jellium~\cite{bernu2008metal,bernu2011hartree,footnote1}.
Related ideas have also appeared in works on Wigner crystals~\cite{fisher1979defects, cockayne1991energetics, candido2001single, nemeth2002ground, katomeris2003andreev, falakshahi2005hybrid, kim_2024_dynamical, han_2019_charge, grover2025critical}, 
as well as in the long-studied ``supersolid'' problem~\cite{leggett1970can, andreev1969quantum, chester1970speculations, kim2004probable, kim2004observation, anderson2005thermodynamics, prokofev2005supersolid,  boninsegni2012colloquium}.

We highlight three central elements of the current work: (I) we propose a general criterion for when commensurate crystals become unstable to self-doping, (II) we compute where instabilities actually occur in a microscopic model of RMG, and (III) we study the properties of the resulting MWC phases and compare to experimental results.

(I)
We first formulate a sufficient condition guaranteeing that the commensurate (insulating) crystal is unstable to doping: the system lowers its energy by spontaneously shifting away from commensurate filling and producing an incommensurate, metallic Wigner crystal.
This condition takes the simple form
\begin{equation}
    \frac{\nu \Delta}{2} < {|k_0|},
\end{equation}
where $\Delta$ is the charge gap of the insulating crystal, $k_0$ is the ``packing bias" of the crystal at commensuration (defined below), and $\nu$ is the filling factor of a unit cell. 
This condition is fully general, and encompasses earlier considerations of interstitials and vacancies~\cite{fisher1979defects,cockayne1991energetics,candido2001single,katomeris2003andreev, kim_2024_dynamical}.

(II) We next apply these considerations to the experimentally relevant setting of RMG under a strong displacement field.
We compute the stability criterion for commensurate crystals across the phase diagram of RMG, which may be done using mean-field solutions of \textit{purely} commensurate crystals. We find that commensurate WCs become unstable to self-doping across large regions of the phase space, and further predict the sign of the MWC charge carriers.

\begin{figure}[h]
    \centering
    \includegraphics[width=\linewidth]{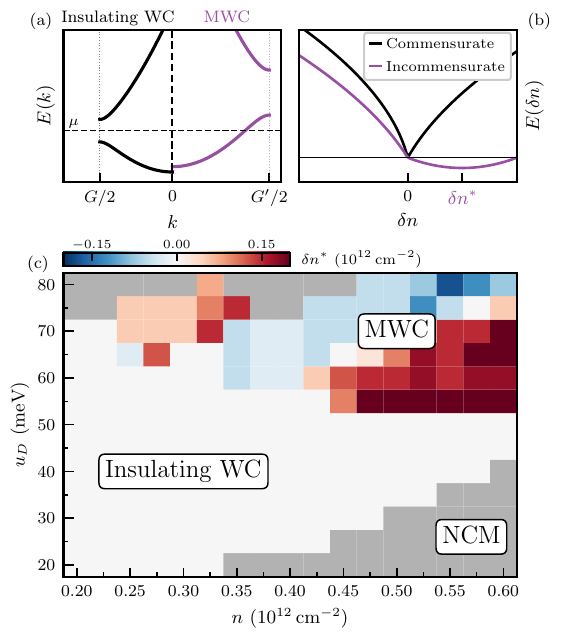}
    \caption{
    (a) Schematic band structures of an insulating Wigner crystal (WC) and a metallic Wigner crystal (MWC), shifted such that at the same chemical potential $\mu$ shown in the plot they correspond to the same electron density. Vertical lines indicate respective crystalline Brillouin zone edges.
    (b) Cartoon dependence of energy on the density deviation defined in Eq.~\eqref{eq:density_deviation}. Insulating WCs generally have a cusp in $E(\delta n)$ at $\delta n=0$ corresponding to its charge gap when pinned. Incommensurate WCs have $\delta n^*\neq 0$.
    (c) Phase diagram of rhombohedral four-layer graphene at low density within SCHF, where crystalline phases including insulating WCs and MWCs compete with the non-crystalline metal (NCM) shown in gray, whose $|\rho_{\v{G}}|A_{uc}<0.05$ for $\v{G}$ in the first shell. The colors show $\delta n^*$, the optimal deviation from commensurate density, defined in Eq.~\eqref{eq:density_deviation}.
    }
    \label{fig:phase_diagram}
\end{figure}

(III) Finally, we carry out self-consistent Hartree-Fock (SCHF) over the landscape of incommensurate crystals, mapping out the phase diagram of optimally doped crystals as a function of carrier density and displacement field. We find that the insulating Wigner crystal is partly superseded by MWC phases that appear directly next to it in parameter space (Fig.~\ref{fig:phase_diagram}).
Recent experiments~\cite{Tonghang_upcoming} in moir\'e-less rhombohedral graphene strikingly observed an island of a metallic phase with reversed sign of the Hall conductance as compared to the neighboring phases (see also Refs.~\cite{han2025signatures, nguyen2025hierarchy}), with unique temperature and magnetic field dependence~\cite{Tonghang_upcoming}. Our calculations provide a natural interpretation of the results in Ref.~\cite{Tonghang_upcoming}. Our hole doped MWC phase appears in a similar density and displacement field range,
which explains the reversed sign of the Hall effect. The calculated hole density is consistent with experiments.  
We further predict the evolution of the metallic Wigner crystal state with density, calculate the collective mode spectrum using time-dependent Hartree-Fock (TDHF) theory, and discuss consistency of our result with  observations in Ref.~\cite{Tonghang_upcoming}.

\PRLsec{Metallic Wigner Crystals} We start by considering a system of $N$ spinless electrons with area $A_s$, {\it i.e.} density $n = N/A_s$,  and continuous translation symmetry. The ground state is \textit{crystalline} if it spontaneously breaks continuous translation symmetry, generating a crystalline lattice with $N_{uc}$ unit cells, each with unit cell area $A_{uc} = A_s / N_{uc}$.

Crystalline states are characterized by their electrical conductivities when weakly pinned. When conductivity vanishes, $\sigma_{xx}=\sigma_{xy}=0$, the state is an insulating Wigner crystal (WC), while states with $\sigma_{xx}=0,\sigma_{xy}\neq 0$ are called Hall crystals. A crystalline state with $\sigma_{xx}\neq 0$ is a \textit{metallic Wigner crystal} (MWC).

Crystalline states are further characterized by their filling $\nu = n/n_{uc}$, where $n_{uc} = A_{uc}^{-1}$. \textit{Commensurate} crystals have $\nu\in\mathbb{Z}$. Commensuration is necessary for insulating crystalline states, but not sufficient since metallic Wigner crystals with equal sized electron and hole pockets are commensurate.
The deviation from commensurability $n/n_{uc} = m\in\mathbb{Z}$ is defined as
\begin{equation}
\label{eq:density_deviation}
      \delta n(n,A_{uc}) = n - mn_{uc} = (\nu-m)A_{uc}^{-1},
\end{equation}
which vanishes at commensuration and is positive (negative) for electron (hole) doping.

\PRLsec{Thermodynamics of crystal commensurability}We now derive a criterion for a commensurate crystal to be stable against incommensurability.

We start by considering an energy functional $E(N, A_s, N_{uc})$~\footnote{We take $N$ and $N_{uc}$ to be continuous variables. Tiling problems at the boundary of the system will generate subextensive energy costs, and thus are of subleading concern.}.
This is a homogeneous function of three extensive variables, which implies
$E(\lambda N, \lambda A_s, \lambda N_{uc}) = \lambda E(N, A_s, N_{uc})$.
Allowing for the possibility of a cusp, thermodynamic stability requires $\lim_{N_{uc}\to N^+}\partial E / \partial N_{uc} \geq  0$, and $\lim_{N_{uc}\to N^-}\partial E / \partial N_{uc} \leq 0$. Due to the homogeneity of $E$, we have
\begin{equation}
    k(N, A_s, N_\text{uc}) \equiv \frac{\partial E}{\partial N_{uc}} = \frac{E - \mu N + \tilde{p} A_s}{N_\text{uc}},
\label{eq:dE/dNuc}
\end{equation}
where $\mu=\partial E/\partial N$ is the chemical potential and $ \tilde{p} = - \partial E/\partial A_s$ is the pressure \textit{computed with $N_{uc}$ held fixed}.
We call $k$ the \textit{packing bias} since its sign indicates whether it is energetically favorable for $N_{uc}$ to increase or decrease~\footnote{
An alternative way to understand $k_0$ is as a Legendre transform of $E$ with respect to $N$ and $A_s$. 
The Legendre transform $N_{uc}k(\mu,\tilde{p},N_{uc})$ is continuous, while $N_\text{uc}k(N,A_s,N_{uc})$ used in the text may be discontinuous.
}.
Defining $k^{\pm}=\lim_{N_{uc} \to N^{\mp}} k(N, A_s, N_{uc})$~\footnote{We define $k^\pm$ such that $k^+$ corresponds to when the system is slightly electron doped, and $k^-$ corresponds to when the system is slightly hole doped.}, the stability condition becomes $k^- \geq 0, k^+ \leq 0$. There are two ways to satisfy this: either $k^+ = k^- = 0$ --- which is unstable to infinitesimal deformations --- or there is a discontinuity $k^- - k^+ > 0$, which provides robust stability.

Insulating crystals are stabilized by a discontinuity in $k$, controlled by their charge gap $\Delta$. As $N_{uc}$ decreases, the crystal goes from hole-doped to electron-doped, creating a chemical potential jump
~\footnote{The charge gap can be nonzero because we hold $N_{uc}$ fixed. If $N_{uc}$ is allowed to vary with $N$, the crystal can breathe to accommodate additional charges at no cost.}
\begin{equation}
    \mu^+ - \mu^- = \Delta,\quad \mu^{\pm}\equiv\lim_{N_{uc} \to N^{\mp}} \mu(N, A_s, N_{uc}).
\end{equation}
This jump is then inherited by the packing bias  $k^{+}-k^{-}=-(\mu^+-\mu^-)N/N_{uc}=-\nu \Delta$~\footnote{We note that $E$ and $\partial E/\partial A_s$ are both continuous functions of $N,A_s,N_{uc}$. To show the second part of the statement, use Eq.~\eqref{eq:eps_n_nuc} and $E=A_s\epsilon$. Since $N,N_{uc}$ are held fixed when taking the partial derivative, the derivative must be continuous at $N = N_{uc}$.}. Defining $k_0 = (k^+ + k^-)/2$, the stability criteria can be concisely stated as
\begin{equation}
    \frac{\nu\Delta}{2} \geq {|k_0|}.
    \label{eq:stability}
\end{equation}
When $k_0$ is large and positive (negative), the crystal is unstable to deformations in which $N_{uc}$ is smaller (larger), thus spontaneously introducing electron-like (hole-like) carriers.

Physically, a commensurate crystal with a larger charge gap is more stable against incommensuration.
Conversely, metallic Wigner crystals with $\Delta=0$ are stabilized where $k^+ = k^- = 0$, making the incommensurate case the generic one; see also Ref.~\cite{Zhihuan_upcoming}. {Previously, a similar argument was proposed in the context of supersolidity in Ref.~\cite{prokofev2005supersolid}.} Furthermore, Eq.~\eqref{eq:stability} may be calculated purely from states on the commensurate line, as no $N_{uc}$ derivatives are required. Below we check it in a microscopic model.

It is instructive to understand the energy landscape in terms of $n$ and $n_{uc}$. In App.~\ref{app:CC} we show that the first order expansion of $\epsilon(n, n_{uc})=E/A_s$ around a commensurate point $n_{uc} = n $ is
\begin{equation}
\begin{aligned}
    \epsilon(n, n_{uc}) &\approx \epsilon_0 + \frac{\Delta}{2} \left| n - n_{uc} \right| + \mu_0 (n - n_0) + k_0(n_{uc}-n_{0}),
\label{eq:eps_n_nuc}
\end{aligned}
\end{equation}
where $\mu_0 = (\mu^+ + \mu^-)/2$. This is smooth along the commensurability line $n = mn_{uc}$, but has a cusp in the $n_{uc}$ direction of size $\Delta$, thereby providing stability to the commensurate crystal. When the cusp is smaller than the threshold in Eq.~\eqref{eq:stability}, the energy density $\epsilon$ can develop minima away from the commensurability line, Fig.~\ref{fig:phase_diagram}(b). This transition may be continuous.

\begin{figure}
    \centering
    \includegraphics[width=\linewidth]{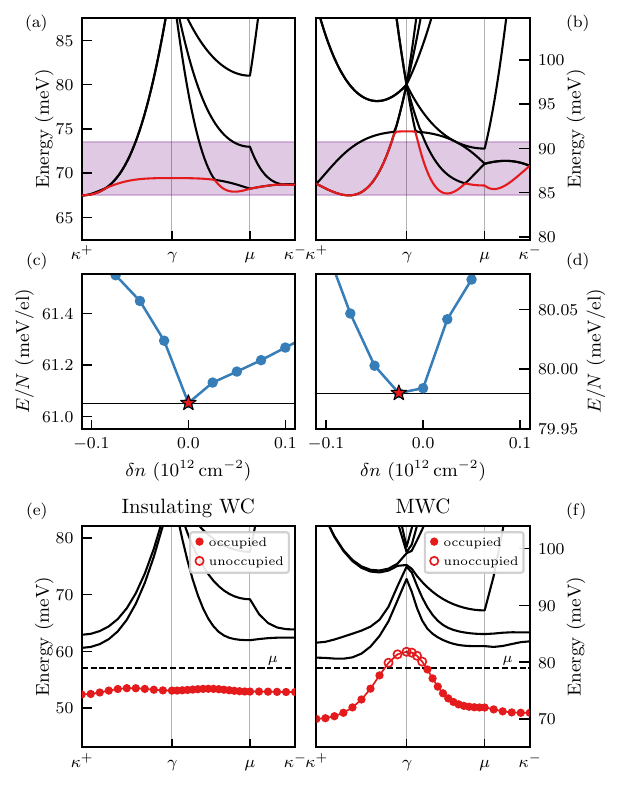}
    \caption{
    Properties of the insulating Wigner crystal at $(n, u_D)=(\SI{0.4e12}{\per\centi\meter\squared},\SI{45}{meV})$ (left) and the metallic Wigner crystal at the same density and $u_D=\SI{60}{meV}$ (right).
    (a,b) Non-interacting dispersion over a high symmetry path of the commensurate crystalline Brillouin zone. The lowest band is highlighted in red. A window of energy corresponding to the interaction strength $U=e^2\sqrt{n}/\epsilon_r\epsilon_0$ is shaded purple.
    (c,d) $E(\delta n)/N$ computed as a function of the density mismatch $\delta n=n-n_{uc}$. The optimal value $\delta n^*$ for which $E$ is minimized is shown with a red star. On the left, $\delta n^*$ is $0$, indicating a commensurate insulating WC, and on the right, $\delta n^*=-\SI{0.025e12}{\per\centi\meter\squared}$, indicating an incommensurate MWC.
    (e,f) SCHF band structures. The lowest band is highlighted in red, with solid circles indicating occupied states and empty circles indicating unoccupied states. On the right, the MWC exhibits a hole pocket at the $\gamma$ point of its crystalline BZ. 
    }
    \label{fig:MWC_properties}
\end{figure}

\PRLsec{Rhombohedral graphene}We now turn to the study of metallic Wigner crystals in rhombohedral graphene. We consider a standard model for interacting rhombohedral multilayer graphene (RMG)
\begin{equation}
    \label{eq:Hamiltonian}
    \hat{H} = \hat{h}_{\mathsf{RMG}} + \frac{1}{2 A_s} \sum_{\v{q}} V_{\v{q}} :\hat{\rho}_{\v{q}} \hat{\rho}_{-\v{q}}:, 
    V_{q} = \frac{2\pi \tanh q d}{\epsilon_r \epsilon_0 q},
\end{equation}
where $\hat{\rho}_{\v{q}}$ is the density operator at momentum $\v{q}$, normal ordering is relative to the charge neutral gap of RMG. The screened Coulomb interactions $V_{q}$ have gate distance
$d=\SI{25}{\nano\meter}$, and relative permittivity $\epsilon_r=15$; the large permittivity is chosen following Ref.~\cite{auerbach2025isospin}, which obtained a good fit with the experimental phase diagram in the same material. Dependence of the phase diagram on the dielectric constant can be found in App.~\ref{app:eps_r_dependence}. The single-particle Hamiltonian $\hat{h}_{\mathsf{RMG}}$ uses a standard tight-binding model for graphene, detailed in App.~\ref{app:RMG}. We model the perpendicular electric field with a potential difference $u_D$ between each adjacent layer.
For concreteness, we focus on four layers, but our results should apply broadly with more layers.

We solve Eq.~\eqref{eq:Hamiltonian} within the self-consistent Hartree-Fock (SCHF) approximation, assuming polarization to a single spin and valley. As we are interested in crystallization, we enforce discrete translation symmetry corresponding to a crystalline lattice $\Lambda = \braket{\v{L}_1,\v{L}_2}$ where $\v{L}_1 = L(1,0)$ and $\v{L}_2 = L (-1/2,\sqrt{3}/2)$. Here $L$ is an adjustable parameter on the $\sim{} \SI{10}{\nano\meter}$ scale, giving a unit cell area $A_{uc} = \frac{\sqrt{3}}{4} L^2$. This also fixes a reciprocal lattice $\Lambda^* =  \braket{\v{G}_1,\v{G}_2}$, and a band structure over its crystalline Brillouin zone. Since we focus on the flat band minima and the low density regime,
we project to the $N_b = 13$ conduction bands closest to the Fermi level. To resolve the small Fermi pockets, large system sizes are required; we use $N_k = 30 \times 30$ unit cells. See App.~\ref{app:RMG} for numerical details. Given that the interparticle spacing is much larger than the graphene lattice constant, the low-energy Hamiltonian has continuous translation symmetry.

Fig.~\ref{fig:phase_diagram}(c) shows the SCHF phase diagram. For each displacement $u_D$ and density $n$, we find the optimal unit cell size by computing the SCHF ground state energy $E(\delta n)$ over a range $|\delta n| \leq \SI{0.2e12}{\per\centi\meter\squared}$ and set $\delta n^* = \operatorname{argmin} E(\delta n)$ (Fig.~\ref{fig:phase_diagram}(b)). To diagnose continuous translation breaking, we examine the order parameter $\braket{\rho_{\v{G}}}$, where $\v{G} \in \Lambda^*$ is a primitive reciprocal lattice vector. We find a non-crystalline metal (NCM) phase as well as crystalline phases. Within the crystalline region, we find that increasing $u_D$ often drives a transition from a Wigner crystal to a metallic Wigner crystal phase, Fig.~\ref{fig:phase_diagram}(c). 
We find both electron- and hole-doped MWC states, and even some perfectly compensated crystals.

\begin{figure}
    \centering
    \includegraphics[width=\linewidth]{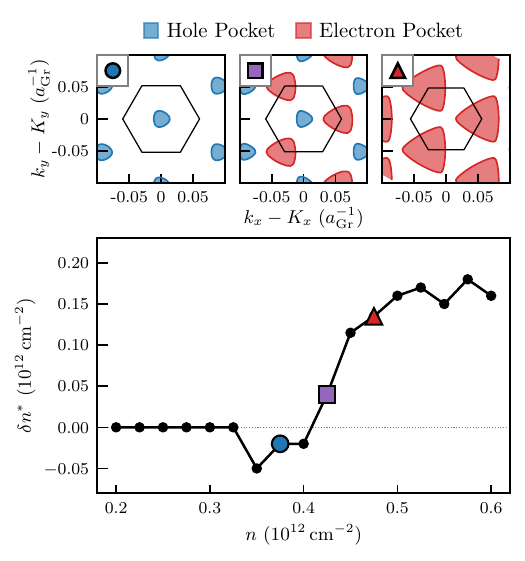}
    \caption{(Top) {Evolving} fermiology of the metallic Wigner crystal at $u_D=\SI{60}{meV}$ and density $n=(0.375, 0.425, 0.475)\times $\SI{e12}{\per\centi\meter\squared} (left to right). Black hexagons show the Brillouin zone corresponding to optimal $n_{uc}=n - \delta n^*$.
    (Bottom) Optimal deviation $\delta n^*$ from commensurate density, Eq.~\eqref{eq:density_deviation}, as a function of density at $u_D=\SI{60}{meV}$.
    }
    \label{fig:Fermi_surfaces}
\end{figure}

\PRLsec{MWC properties} 
We now investigate the properties of the MWC phase. Fig.~\ref{fig:MWC_properties} 
compares and contrasts the WC and MWC by focusing on two representative points, $(n,u_D)=(\SI{0.4 e12}{\per\centi\meter\squared}, \SI{45}{meV})$ on the left and $(n,u_D)=(\SI{0.4 e12}{\per\centi\meter\squared}, \SI{60}{meV})$ on the right. 
The single-particle dispersion of RMG has a ``Mexican hat" profile, 
which is almost flat near $u_D = \SI{45}{meV}$ but develops a central peak at higher $u_D$. This is backfolded by the crystal lattice, creating a non-interacting miniband with maximum at $\gamma$ and minimum near the zone boundary in this regime, whose separation can exceed the interaction scale $U=e^2\sqrt{n}/\epsilon_r\epsilon_0$ at large $u_D$, Fig.~\ref{fig:MWC_properties}(a,b).

Fig.~\ref{fig:MWC_properties}(c,d) shows the energy within SCHF as a function of $\delta n$ at both points. In (c) the energy has a cusp with minimum at $\delta n^*=0$, indicating an insulating WC, while in (d) the energy is minimized at incommensurate density $\delta n^*=\SI{-0.025 e12}{\per\centi\meter\squared}$, giving an MWC. Fig.~\ref{fig:MWC_properties}(e,f) show their SCHF band structures. The WC has a charge gap of $O(\SI{10}{meV})$ above the first band, whereas the lowest MWC band is peaked at $\gamma$, creating a small hole pocket (f). 
The holes at the Fermi surface have effective mass of ${\sim}0.05 m_e$, significantly lighter than their non-interacting value. 
Below we will explain how the changing band structure helps drive the WC to MWC transition.

Fig.~\ref{fig:Fermi_surfaces} shows how the Fermi surface of MWCs evolves with density at 
$u_D=60\si{meV}$. As density evolves, the WC first develops small hole pockets, then becomes (imperfectly) compensated, and finally develops large electron pockets. Their placement is consistent: hole pockets appear at $\gamma$, while electron pockets are always centered at $\kappa^+$.

\begin{figure}
    \centering
    \includegraphics[width=\linewidth]{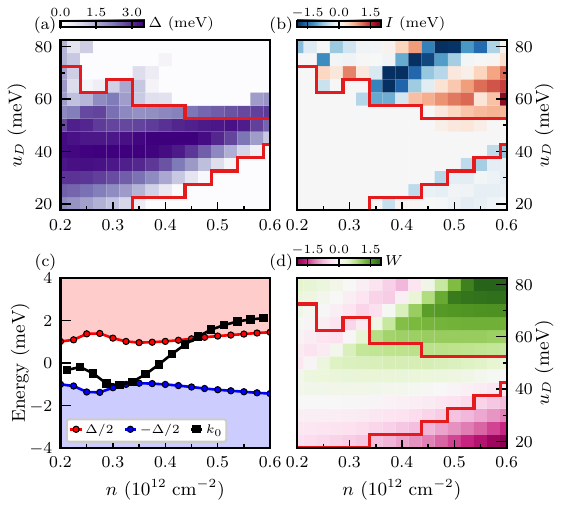}
    \caption{
    (a) The charge gap $\Delta$ of the SCHF ground state at commensurate filling $\delta n = 0$ as a function of $(n,u_D)$.
    A smaller charge gap at commensuration leads to either MWCs or NCMs. The phase boundary corresponding to the commensurate insulating WC is shown in red.
    (b) The instability $I$, Eq.~\eqref{eq:instability}, for the commensurate SCHF state as a function of $(n,u_D)$. A positive (negative) sign of $I$ signals instability to electron (hole) doping.
    (c) Comparison of packing bias $k_0$ and $\pm \Delta /2$ as a function of density at $u_D = \SI{55}{meV}$. $k_0$ changes rapidly while $\pm \Delta/2$ remains roughly constant in $n$, illuminating the delicate competition between $\Delta$ and $k_0$.
    (d) Map of the single particle heuristic $W=(E_{\gamma}-E_{\kappa^+})/U$, where $E_k$ corresponds to the lowest band after folding the non-interacting band structure on to a commensurate crystalline Brillouin zone and $U$ is the interaction scale $e^2\sqrt{n}/\epsilon_r\epsilon_0$. Large $W$ suggests instability towards an MWC.
    }
    \label{fig:CC_heuristic}
\end{figure}

\PRLsec{Instability of commensurate Wigner crystal}
The appearance of MWCs can be predicted by perturbative instability of commensurate Wigner crystal controlled by $\Delta$ and $k_0$.  To assess this, Fig.~\ref{fig:CC_heuristic}(a) plots the charge gap $\Delta$ at commensurate filling $\delta n = 0$ as a function of $(n,u_D)$. This shows that $\Delta$ is large inside the insulating WC phase, whose phase boundary from Fig.~\ref{fig:phase_diagram}(c) is shown here as a red line, while NCM and MWC states appear in regions with small $\Delta$. To check the tendency towards electron or hole doping, we define the \textit{instability}
\begin{equation}
    I = \begin{cases}
        k_0  + \frac{\nu \Delta}{2}&\text{if }  k_0 < -\frac{\nu\Delta}{2} \text{ (unstable to holes)}
        \\[0.3em]
        k_0 - \frac{\nu\Delta}{2}&\text{if }  k_0 > \frac{\nu\Delta}{2}
        \text{ (unstable to electrons)}
        \\
        0 & \text{otherwise},
    \end{cases}
    \label{eq:instability}
\end{equation}
which is chosen so the sign of $I$ predicts the sign of $\delta n^*$. This is shown in Fig.~\ref{fig:CC_heuristic}(b), which accurately describes the sign of the charge carrier in Fig.~\ref{fig:phase_diagram}(c). Plotting $k_0$ and $\pm \Delta/2$ in Fig.~\ref{fig:CC_heuristic}(c) reveals that the stability is often governed by a delicate balance between these quantities, which we analyze further in App.~\ref{app:analysis_stability}.

\PRLsec{Single-particle heuristic}
We now provide a simple heuristic to predict where the MWC will appear in RMG by estimating $\Delta$ from the single particle bandstructure. The ``Mexican hat" profile of RMG creates a local maximum at $\gamma$, and a minimum in the vicinity of the triply-degenerate $\kappa^+$ point. In a weak-coupling picture, interactions will open a charge gap of order the interaction strength $U = e^2 \sqrt{n} /\epsilon_r \epsilon_0$ at the zone boundaries. A charge gap $\Delta$ can only result if interactions are strong enough to push $\kappa^+$ of the second band above the $\gamma$ maximum of the first band, 
which requires the single particle energies to obey $U \gtrsim E_{\gamma} - E_{\kappa^+}$. Therefore, a large $\Delta$ is favored when the magnitude of
\begin{equation}
    W = (E_\gamma - E_{\kappa^+}) / U
\end{equation}
is small, while a large and positive $W$ suggests an MWC phase.

Fig.~\ref{fig:CC_heuristic}(d) shows $W$ over the $(n,u_D)$ plane. We find that the insulating Wigner crystal phase indeed appears where $W$ is close to zero (white), while the MWC phase matches well with the regime of large and positive $W$ (green). A large region of negative $W$ (pink) coincides with the NCM phase at large density and low displacement field.
However, $W$ is only sensitive to $\Delta$ and not $k_0$, making it unable to predict the carrier type of the MWC. Nevertheless, we conclude $W$ is a good heuristic indicator of instability of a WC to an MWC.

\begin{figure}[htbp]
    \centering
    \includegraphics[width=\linewidth]{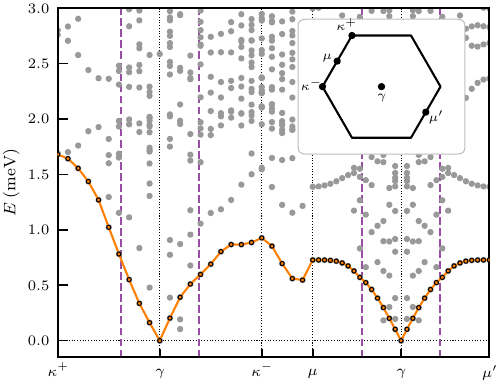}
    \caption{
    Time-dependent Hartree-Fock (TDHF) spectrum of the metallic Wigner crystal. 
    The orange line denotes the acoustic phonon branch.
    Purple dashed lines mark $2k_F$ for a circular Fermi surface of the corresponding density.
    Parameters match Fig.~\ref{fig:MWC_properties}(f).
    }
    \label{fig:TDHF_MWC}
\end{figure}

\PRLsec{TDHF}
Fig.~\ref{fig:TDHF_MWC} shows the spectrum of low-energy neutral excitations above the ground state of the MWC from Fig.~\ref{fig:MWC_properties}(f), computed using time-dependent Hartree-Fock (TDHF).
All eigenvalues have negligible imaginary parts, indicating the perturbative stability of the SCHF ground state. The low-energy spectrum has features from both Goldstone modes (phonons) due to breaking continuous translation symmetry and particle-hole excitations close to the Fermi surface. 
Even at this large ($30 \times 30$) system size, finite size errors in the continuum are of order $\hbar v_F \delta k \approx \SI{1}{meV}$; in the thermodynamic limit the particle-hole continuum should extend to zero energy for all momentum transfers that connect two points on the Fermi surface. We conjecture that the electron-phonon coupling may be significant, since the phonons of electron crystals consist of particle-hole excitations --- identical in nature to the particle-hole continuum --- unlike atomic phonons which are characteristically distinct. However, the estimation of electron-phonon coupling is unfeasible at current momentum resolution due to finite size effects and will be left for future work.

The TDHF spectrum shows the speed of sound $v_s$ is highly directional.
Along $\gamma \to \kappa^+$ we estimate $v_s \approx \SI{9.7}{km/s}$, while from $\gamma \to \kappa^-$ we find $v_s \approx \SI{12.2}{km/s}$, roughly a $25\%$ difference. This indicates a large kineoelastic term, which is generically allowed in the absence of inversion and time-reversal~\cite{AHC4}. Representative TDHF spectrum for a Wigner crystal state can be found in Fig.~\ref{fig:TDHF_WC}, which hosts an even larger kineoelastic term.

\PRLsec{Implications and conjectures}
We conclude with implications of our results on the interpretation of experimental results and guidance on future experiments.

Experimentally, a conducting hole pocket with high mobility was observed in four layer RMG near $n\approx \SI{0.3e12}{\per\centi\meter\squared}$ when the displacement field is increased above the presumed Wigner crystal phase boundary~\cite{Tonghang_upcoming}. (Similar results are found in five and six layer RMG~\cite{Tonghang_upcoming}.) Applying magnetic fields as weak as around $\SI{1}{T}$ creates integer quantized Hall plateaus across much of the pocket, corresponding to hole densities around $\delta n \approx -\SI{0.03e12}{\per\centi\meter\squared}$.
Unusually, at low temperatures, the Hall conductance in the pocket at fixed magnetic field is {\em opposite} to that of the surrounding phases, consistent with previous observations in~\cite{han2025signatures,nguyen2025hierarchy}, but reverses sign to agree with the neighboring phases as temperature is increased.
We posit that these phenomena are naturally explained by the hole-doped MWC found in our phase diagram (Fig.~\ref{fig:phase_diagram}), which exhibits a hole pocket of size $\delta n^* = \SI{-0.025e12}{\per\centi\meter\squared}$ near $u_D\approx \SI{60}{meV}$ and at comparable densities $n\approx \SI{0.4e12}{\per\centi\meter\squared}$.
In a weak magnetic field, we expect the light carriers of the hole pocket to quickly form Landau levels and exhibit quantized Hall conductance --- with a reversed sign relative to the large electron pockets found in the single-flavor metal (``quarter metal")~\cite{zhou2021half, footnote2}. 
Upon raising temperature, the crystalline order is expected to melt, and the hole pocket will transform into the large electron Fermi surface of the single-flavor metal. This accounts for the temperature dependence of Hall conductance seen in the experiment.

We further predict that, as $n$ increases at constant displacement field, the system will remain in the MWC phase, undergoing the same sequence of fermiologies
from Fig.~\ref{fig:Fermi_surfaces}: hole-doped, then (imperfectly) compensated, then finally electron doped. In the compensated MWC, transport and magnetotransport will have contributions from both the hole and electron pockets. The electron pocket has an effective mass approximately $5$ times larger than the hole pocket, suggesting that 
it will be quantized and become partial Hall crystals~\cite{tesanovic_hall_1989} only in larger magnetic fields. Eq.~\eqref{eq:instability} implies that these partial Hall crystals are less stable due to a smaller charge gap, which is set by the smaller of the Landau level gaps from the electron and hole pockets. This may be related to observations of re-entrant integer quantum Hall states in Ref.~\cite{nguyen2025hierarchy}. The complex interplay between crystallization, magnetic fields, and fermiology here will be the topic of future work.

Signatures of MWCs should be measurable in a range of experimental probes beyond transport. The Fermi surface and crystalline order, which backfolds the bands due to Bragg scattering, manifest in the momentum-resolved spectral function, making it accessible to probes such as the quantum twisting microscope~\cite{QTM}. The collective excitations of the MWC phase may be accessible through AC conductivity or surface acoustic waves as low-energy resonances. The coexistence of a crystal and a metal has the advantage that its crystalline collective modes may be excited through their coupling to the Fermi surface of the metal. Finally, scanning tunneling microscopy (STM) would be a natural probe for the MWC phase, albeit a challenging one, due to the large displacement field required. STM would be able to resolve both the crystalline order via charge density variations but also the finite density of states at the Fermi energy.

MWCs can become more favorable than WC at finite temperature due to entropic considerations. Both phases host phonons, which gives rise to a $T^2$ entropy contribution. However, the Fermi surface in MWCs contributes entropy proportional to $T$, dominating the phonons at low temperatures. We note that the correction of entropy due to defect configurations was proposed to be $T^5$~\cite{anderson2005thermodynamics}, which is subleading at low temperatures. This entropic effect tends to expand MWC into the WC region at finite temperature. These considerations are only valid when the crystalline order has not melted. Using the zero-temperature shear modulus extracted from the TDHF spectrum, we estimate the transition temperature to be approximately $\SI{250}{mK}$~\cite{AHC4}. This value is, however, sensitive to model details such as the dielectric constant.

Commensurate anomalous Hall crystals (AHCs) have been identified in the phase diagram of several interacting models with nontrivial Berry curvature including moir\'e-less rhombohedral graphene at larger fillings. Our results suggest that in portions of this phase diagram the commensurate AHC phase may be unstable to self doping, leading to an MWC~\footnote{We indeed confirm that this is the case for certain parameters in the phase diagram.}. This could explain why AHCs remain to be seen experimentally, and a more complete scan of the phase diagram will help identify experimental knobs (density, displacement field, dielectric screening, among others) that could stabilize the commensurate AHC in rhombohedral multilayer graphene.

Finally, given that the insulator and metals discussed here arise from Wigner crystals, 
it is natural to conjecture that the chiral superconductor found in Ref.~\cite{han2025signatures} also has crystalline order ({\it i.e.} a supersolid), and owes its existence to the phonons of the crystal. The potential implications of this scenario are left for future work.

\PRLsec{Note added}
Related results with a partial overlap with present work will appear in a forthcoming preprint~\cite{Zhihuan_upcoming}.

\PRLsec{Acknowledgements}
The authors are particularly thankful to Tonghang Han and Long Ju for many fruitful discussions, and for sharing their data prior to publication of Ref.~\cite{Tonghang_upcoming}. We would like to thank Zhihuan Dong, Jiechao Feng, Zhaoyu Han, Michael P. Zaletel, Aaron Sharpe, Felix Desrochers, and Tarun Grover for helpful discussions.
A.V. and J.D. acknowledge funding from NSF DMR-2220703 and the Simons Collaboration on Ultra-Quantum Matter, which is a grant from the Simons Foundation (651440, A.V.). 
D.E.P. acknowledges support from NSF CAREER award no. DMR-2542485.
The Flatiron Institute is a division of the Simons Foundation.


\bibliographystyle{unsrt}
\bibliography{references}

\begin{thebibliography}{10}

\bibitem{wigner_interaction_1934}
E.~Wigner.
\newblock On the {Interaction} of {Electrons} in {Metals}.
\newblock {\em Phys. Rev.}, 46(11):1002--1011, December 1934.
\newblock Publisher: American Physical Society.

\bibitem{Giuliani_Vignale_2005}
Gabriele Giuliani and Giovanni Vignale.
\newblock {\em Quantum Theory of the Electron Liquid}.
\newblock Cambridge University Press, 2005.

\bibitem{tanatar1989ground}
B~Tanatar and David~M Ceperley.
\newblock Ground state of the two-dimensional electron gas.
\newblock {\em Physical Review B}, 39(8):5005, 1989.

\bibitem{Drummond_Phase_Diagram}
N.~D. Drummond and R.~J. Needs.
\newblock Phase diagram of the low-density two-dimensional homogeneous electron gas.
\newblock {\em Phys. Rev. Lett.}, 102:126402, Mar 2009.

\bibitem{lozovik_crystallization_1975}
Yu.~E. Lozovik and V.~I. Yudson.
\newblock Crystallization of a two-dimensional electron gas in a magnetic field.
\newblock {\em Soviet Journal of Experimental and Theoretical Physics Letters}, 22:11, July 1975.
\newblock ADS Bibcode: 1975JETPL..22...11L.

\bibitem{grimes_evidence_1979}
C.~C. Grimes and G.~Adams.
\newblock Evidence for a {Liquid}-to-{Crystal} {Phase} {Transition} in a {Classical}, {Two}-{Dimensional} {Sheet} of {Electrons}.
\newblock {\em Phys. Rev. Lett.}, 42(12):795--798, March 1979.
\newblock Publisher: American Physical Society.

\bibitem{andrei_observation_1988}
E.~Y. Andrei, G.~Deville, D.~C. Glattli, F.~I.~B. Williams, E.~Paris, and B.~Etienne.
\newblock Observation of a {Magnetically} {Induced} {Wigner} {Solid}.
\newblock {\em Phys. Rev. Lett.}, 60(26):2765--2768, June 1988.
\newblock Publisher: American Physical Society.

\bibitem{yoon_wigner_1999}
Jongsoo Yoon, C.~C. Li, D.~Shahar, D.~C. Tsui, and M.~Shayegan.
\newblock Wigner {Crystallization} and {Metal}-{Insulator} {Transition} of {Two}-{Dimensional} {Holes} in {GaAs} at \${B}=0\$.
\newblock {\em Phys. Rev. Lett.}, 82(8):1744--1747, February 1999.
\newblock Publisher: American Physical Society.

\bibitem{ma_thermal_2020}
Meng~K. Ma, K.~A. Villegas~Rosales, H.~Deng, Y.~J. Chung, L.~N. Pfeiffer, K.~W. West, K.~W. Baldwin, R.~Winkler, and M.~Shayegan.
\newblock Thermal and {Quantum} {Melting} {Phase} {Diagrams} for a {Magnetic}-{Field}-{Induced} {Wigner} {Solid}.
\newblock {\em Phys. Rev. Lett.}, 125(3):036601, July 2020.
\newblock Publisher: American Physical Society.

\bibitem{smolenski_signatures_2021}
Tomasz Smoleński, Pavel~E. Dolgirev, Clemens Kuhlenkamp, Alexander Popert, Yuya Shimazaki, Patrick Back, Xiaobo Lu, Martin Kroner, Kenji Watanabe, Takashi Taniguchi, Ilya Esterlis, Eugene Demler, and Ata{\oldc c} Imamo{\oldv g}lu.
\newblock Signatures of {Wigner} crystal of electrons in a monolayer semiconductor.
\newblock {\em Nature}, 595(7865):53--57, July 2021.
\newblock Publisher: Nature Publishing Group.

\bibitem{zhou_bilayer_2021}
You Zhou, Jiho Sung, Elise Brutschea, Ilya Esterlis, Yao Wang, Giovanni Scuri, Ryan~J. Gelly, Hoseok Heo, Takashi Taniguchi, Kenji Watanabe, Gergely Zaránd, Mikhail~D. Lukin, Philip Kim, Eugene Demler, and Hongkun Park.
\newblock Bilayer {Wigner} crystals in a transition metal dichalcogenide heterostructure.
\newblock {\em Nature}, 595(7865):48--52, July 2021.
\newblock Publisher: Nature Publishing Group.

\bibitem{yang_experimental_2021}
Fangyuan Yang, Alexander~A. Zibrov, Ruiheng Bai, Takashi Taniguchi, Kenji Watanabe, Michael~P. Zaletel, and Andrea~F. Young.
\newblock Experimental {Determination} of the {Energy} per {Particle} in {Partially} {Filled} {Landau} {Levels}.
\newblock {\em Phys. Rev. Lett.}, 126(15):156802, April 2021.
\newblock Publisher: American Physical Society.

\bibitem{falson_competing_2022}
J.~Falson, I.~Sodemann, B.~Skinner, D.~Tabrea, Y.~Kozuka, A.~Tsukazaki, M.~Kawasaki, K.~von Klitzing, and J.~H. Smet.
\newblock Competing correlated states around the zero-field {Wigner} crystallization transition of electrons in two dimensions.
\newblock {\em Nat. Mater.}, 21(3):311--316, March 2022.
\newblock Publisher: Nature Publishing Group.

\bibitem{sung_observation_2023}
Jiho Sung, Jue Wang, Ilya Esterlis, Pavel~A. Volkov, Giovanni Scuri, You Zhou, Elise Brutschea, Takashi Taniguchi, Kenji Watanabe, Yubo Yang, Miguel~A. Morales, Shiwei Zhang, Andrew~J. Millis, Mikhail~D. Lukin, Philip Kim, Eugene Demler, and Hongkun Park.
\newblock Observation of an electronic microemulsion phase emerging from a quantum crystal-to-liquid transition, December 2023.
\newblock arXiv:2311.18069 [cond-mat].

\bibitem{tsui_direct_2023}
Yen-Chen Tsui, Minhao He, Yuwen Hu, Ethan Lake, Taige Wang, Kenji Watanabe, Takashi Taniguchi, Michael~P. Zaletel, and Ali Yazdani.
\newblock Direct observation of a magnetic field-induced {Wigner} crystal, December 2023.
\newblock arXiv:2312.11632 [cond-mat].

\bibitem{xiang_quantum_2024}
Ziyu Xiang, Hongyuan Li, Jianghan Xiao, Mit~H. Naik, Zhehao Ge, Zehao He, Sudi Chen, Jiahui Nie, Shiyu Li, Yifan Jiang, Renee Sailus, Rounak Banerjee, Takashi Taniguchi, Kenji Watanabe, Sefaattin Tongay, Steven~G. Louie, Michael~F. Crommie, and Feng Wang.
\newblock Quantum {Melting} of a {Disordered} {Wigner} {Solid}, February 2024.
\newblock arXiv:2402.05456 [cond-mat].

\bibitem{zhou2025electroniccrystalslayeredmaterials}
You Zhou, Ilya Esterlis, and Tomasz Smoleński.
\newblock Electronic crystals in layered materials, 2025.

\bibitem{zhou2021half}
Haoxin Zhou, Tian Xie, Areg Ghazaryan, Tobias Holder, James~R Ehrets, Eric~M Spanton, Takashi Taniguchi, Kenji Watanabe, Erez Berg, Maksym Serbyn, et~al.
\newblock Half-and quarter-metals in rhombohedral trilayer graphene.
\newblock {\em Nature}, 598(7881):429--433, 2021.

\bibitem{hallcrystal2022}
Anna~M. Seiler, Fabian~R. Geisenhof, Felix Winterer, Kenji Watanabe, Takashi Taniguchi, Tianyi Xu, Fan Zhang, and R.~Thomas Weitz.
\newblock Quantum cascade of correlated phases in trigonally warped bilayer graphene.
\newblock {\em Nature}, 608(7922):298--302, August 2022.

\bibitem{lu2024fractional}
Zhengguang Lu, Tonghang Han, Yuxuan Yao, Aidan~P Reddy, Jixiang Yang, Junseok Seo, Kenji Watanabe, Takashi Taniguchi, Liang Fu, and Long Ju.
\newblock Fractional quantum anomalous hall effect in multilayer graphene.
\newblock {\em Nature}, 626(8000):759--764, 2024.

\bibitem{waters2025chern}
Dacen Waters, Anna Okounkova, Ruiheng Su, Boran Zhou, Jiang Yao, Kenji Watanabe, Takashi Taniguchi, Xiaodong Xu, Ya-Hui Zhang, Joshua Folk, et~al.
\newblock Chern insulators at integer and fractional filling in moir{\'e} pentalayer graphene.
\newblock {\em Physical Review X}, 15(1):011045, 2025.

\bibitem{xiang2025continuouslytunableanomaloushall}
Hanxiao Xiang, Jing Ding, Jiannan Hua, Naitian Liu, Wenqiang Zhou, Qianmei Chen, Kenji Watanabe, Takashi Taniguchi, Na~Xin, Wei Zhu, and Shuigang Xu.
\newblock Continuously tunable anomalous hall crystals in rhombohedral heptalayer graphene, 2025.

\bibitem{zhang2010band}
Fan Zhang, Bhagawan Sahu, Hongki Min, and Allan~H MacDonald.
\newblock Band structure of a b c-stacked graphene trilayers.
\newblock {\em Physical Review B}, 82(3):035409, 2010.

\bibitem{AHC3}
Tomohiro Soejima, Junkai Dong, Ashvin Vishwanath, and Daniel~E. Parker.
\newblock $\ensuremath{\lambda}$-jellium model for the anomalous hall crystal.
\newblock {\em Phys. Rev. Lett.}, 135:186505, Oct 2025.

\bibitem{Tan_parent_berry}
Tixuan Tan and Trithep Devakul.
\newblock Parent berry curvature and the ideal anomalous hall crystal.
\newblock {\em Phys. Rev. X}, 14:041040, Nov 2024.

\bibitem{valenti2025quantumgeometrydrivencrystallization}
Agnes Valenti, Yaar Vituri, Yubo Yang, Daniel~E. Parker, Tomohiro Soejima, Junkai Dong, Miguel~A. Morales, Ashvin Vishwanath, Erez Berg, and Shiwei Zhang.
\newblock Quantum geometry driven crystallization: A neural-network variational monte carlo study, 2025.

\bibitem{joy2025chiral}
Sandeep Joy, Leonid Levitov, and Brian Skinner.
\newblock Chiral wigner crystal phases induced by berry curvature.
\newblock {\em Phys. Rev. Lett.}, 135:256502, Dec 2025.

\bibitem{AHC_Yahui}
Boran Zhou, Hui Yang, and Ya-Hui Zhang.
\newblock Fractional quantum anomalous hall effect in rhombohedral multilayer graphene in the moir\'eless limit.
\newblock {\em Phys. Rev. Lett.}, 133:206504, Nov 2024.

\bibitem{AHC1}
Junkai Dong, Taige Wang, Tianle Wang, Tomohiro Soejima, Michael~P. Zaletel, Ashvin Vishwanath, and Daniel~E. Parker.
\newblock Anomalous hall crystals in rhombohedral multilayer graphene. i. interaction-driven chern bands and fractional quantum hall states at zero magnetic field.
\newblock {\em Phys. Rev. Lett.}, 133:206503, Nov 2024.

\bibitem{kivelson_cooperative_1986}
Steven Kivelson, C.~Kallin, Daniel~P. Arovas, and J.~R. Schrieffer.
\newblock Cooperative ring exchange theory of the fractional quantized {Hall} effect.
\newblock {\em Phys. Rev. Lett.}, 56(8):873--876, February 1986.
\newblock Publisher: American Physical Society.

\bibitem{halperin_compatibility_1986}
B.~I. Halperin, Z.~Te\oldv{s}anovi\'c, and F.~Axel.
\newblock Compatibility of {Crystalline} {Order} and the {Quantized} {Hall} {Effect}.
\newblock {\em Phys. Rev. Lett.}, 57(7):922--922, August 1986.
\newblock Publisher: American Physical Society.

\bibitem{kivelson_cooperative_1987}
Steven Kivelson, C.~Kallin, Daniel~P. Arovas, and J.~Robert Schrieffer.
\newblock Cooperative ring exchange and the fractional quantum {Hall} effect.
\newblock {\em Phys. Rev. B}, 36(3):1620--1646, July 1987.
\newblock Publisher: American Physical Society.

\bibitem{tesanovic_hall_1989}
Zlatko Te{\oldv{s}}anovi{\'c}, Fran{\oldc{c}}oise Axel, and B.~I. Halperin.
\newblock ``{Hall} crystal'' versus {Wigner} crystal.
\newblock {\em Phys. Rev. B}, 39(12):8525--8551, April 1989.
\newblock Publisher: American Physical Society.

\bibitem{AHC2}
Tomohiro Soejima, Junkai Dong, Taige Wang, Tianle Wang, Michael~P. Zaletel, Ashvin Vishwanath, and Daniel~E. Parker.
\newblock Anomalous hall crystals in rhombohedral multilayer graphene. ii. general mechanism and a minimal model.
\newblock {\em Phys. Rev. B}, 110:205124, Nov 2024.

\bibitem{Zhihuan_Stability}
Zhihuan Dong, Adarsh~S. Patri, and T.~Senthil.
\newblock Stability of anomalous hall crystals in multilayer rhombohedral graphene.
\newblock {\em Phys. Rev. B}, 110:205130, Nov 2024.

\bibitem{zhou2024newclassesquantumanomalous}
Boran Zhou and Ya-Hui Zhang.
\newblock New classes of quantum anomalous hall crystals in multilayer graphene, 2024.

\bibitem{guo2025correlationstabilizedanomaloushall}
Zhongqing Guo and Jianpeng Liu.
\newblock Correlation stabilized anomalous hall crystal in bilayer graphene, 2025.

\bibitem{crepel2025efficient}
Valentin Cr{\'e}pel and Jennifer Cano.
\newblock Efficient prediction of superlattice and anomalous miniband topology from quantum geometry.
\newblock {\em Physical Review X}, 15(1):011004, 2025.

\bibitem{bernevig2025berrytrashcanmodelinteracting}
B.~Andrei Bernevig and Yves~H. Kwan.
\newblock "berry trashcan" model of interacting electrons in rhombohedral graphene, 2025.

\bibitem{Zeng_sublattice_structure}
Yongxin Zeng, Daniele Guerci, Valentin Cr\'epel, Andrew~J. Millis, and Jennifer Cano.
\newblock Sublattice structure and topology in spontaneously crystallized electronic states.
\newblock {\em Phys. Rev. Lett.}, 132:236601, Jun 2024.

\bibitem{Tan_FAHC}
Tixuan Tan, Julian May-Mann, and Trithep Devakul.
\newblock Wavefunction approach to the fractional anomalous hall crystal, 2024.

\bibitem{Zeng_sliding}
Yongxin Zeng and Andrew~J. Millis.
\newblock Berry phase dynamics of sliding electron crystals, 2024.

\bibitem{AHC4}
Junkai Dong, Ophelia~Evelyn Sommer, Tomohiro Soejima, Daniel~E Parker, and Ashvin Vishwanath.
\newblock Phonons in electron crystals with berry curvature.
\newblock {\em Proceedings of the National Academy of Sciences}, 122(38):e2515532122, 2025.

\bibitem{desrochers2025elasticresponseinstabilitiesanomalous}
Félix Desrochers, Mark~R. Hirsbrunner, Joe Huxford, Adarsh~S. Patri, T.~Senthil, and Yong~Baek Kim.
\newblock Elastic response and instabilities of anomalous hall crystals, 2025.

\bibitem{patri2025family}
Adarsh~S. Patri and Marcel Franz.
\newblock Family of multilayer graphene superconductors with tunable chirality: Momentum-space vortices nucleated by a ring of berry curvature.
\newblock {\em Physical Review B}, 112(21), December 2025.

\bibitem{lu2026generalmanybodyperturbationframework}
Xin Lu, Yuanfan Yang, Zhongqing Guo, and Jianpeng Liu.
\newblock General many-body perturbation framework for moir\'e systems, 2026.

\bibitem{desrochers2026electronic}
F\'elix Desrochers, Joe Huxford, Mark~R. Hirsbrunner, and Yong~Baek Kim.
\newblock Electronic crystal phases in the presence of nonuniform berry curvature and tunable berry flux: The ${\ensuremath{\lambda}}_{N}$-jellium model.
\newblock {\em Phys. Rev. B}, 113:045148, Jan 2026.

\bibitem{tan2025ideallimitrhombohedralgraphene}
Tixuan Tan, Patrick~J. Ledwith, and Trithep Devakul.
\newblock The ideal limit of rhombohedral graphene: Interaction-induced layer-skyrmion lattices and their collective excitations, 2025.

\bibitem{overhauser1960giant}
AW~Overhauser.
\newblock Giant spin density waves.
\newblock {\em Physical Review Letters}, 4(9):462, 1960.

\bibitem{overhauser1962spin}
AW~Overhauser.
\newblock Spin density waves in an electron gas.
\newblock {\em Physical Review}, 128(3):1437, 1962.

\bibitem{bernu2008metal}
B~Bernu, F~Delyon, M~Duneau, and M~Holzmann.
\newblock Metal-insulator transition in the hartree-fock phase diagram of the fully polarized homogeneous electron gas in two dimensions.
\newblock {\em Physical Review B—Condensed Matter and Materials Physics}, 78(24):245110, 2008.

\bibitem{bernu2011hartree}
B~Bernu, F~Delyon, M~Holzmann, and L~Baguet.
\newblock Hartree-fock phase diagram of the two-dimensional electron gas.
\newblock {\em Physical Review B—Condensed Matter and Materials Physics}, 84(11):115115, 2011.

\bibitem{footnote1}
We note, however, these mean-field states seem to give way to translation-invariant metals once quantum fluctuation effects are taken into account~\cite{tanatar1989ground, Drummond_Phase_Diagram}.

\bibitem{fisher1979defects}
Daniel~S Fisher, BI~Halperin, and R~Morf.
\newblock Defects in the two-dimensional electron solid and implications for melting.
\newblock {\em Physical Review B}, 20(11):4692, 1979.

\bibitem{cockayne1991energetics}
Eric Cockayne and Veit Elser.
\newblock Energetics of point defects in the two-dimensional wigner crystal.
\newblock {\em Physical Review B}, 43(1):623, 1991.

\bibitem{candido2001single}
Ladir C{\^a}ndido, Philip Phillips, and DM~Ceperley.
\newblock Single and paired point defects in a 2d wigner crystal.
\newblock {\em Physical review letters}, 86(3):492, 2001.

\bibitem{nemeth2002ground}
Z{\'A}~N{\'e}meth and J-L Pichard.
\newblock Ground state of a partially melted wigner molecule.
\newblock {\em EPL (Europhysics Letters)}, 58(5):744--750, 2002.

\bibitem{katomeris2003andreev}
Georgios Katomeris, Franck Selva, and J-L Pichard.
\newblock Andreev-lifshitz supersolid revisited for a few electrons on a square lattice. i.
\newblock {\em The European Physical Journal B-Condensed Matter and Complex Systems}, 31(3):401--412, 2003.

\bibitem{falakshahi2005hybrid}
Houman Falakshahi and Xavier Waintal.
\newblock Hybrid phase at the quantum melting of the wigner crystal.
\newblock {\em Phys. Rev. Lett.}, 94:046801, Jan 2005.

\bibitem{kim_2024_dynamical}
Kyung-Su Kim, Ilya Esterlis, Chaitanya Murthy, and Steven~A. Kivelson.
\newblock Dynamical defects in a two-dimensional wigner crystal: Self-doping and kinetic magnetism.
\newblock {\em Phys. Rev. B}, 109:235130, Jun 2024.

\bibitem{han_2019_charge}
Zhaoyu Han, Shiwei Zhang, and Xi~Dai.
\newblock Charge density waves in a quantum plasma.
\newblock {\em Phys. Rev. B}, 100:155132, Oct 2019.

\bibitem{grover2025critical}
Tarun Grover and John McGreevy.
\newblock A critical theory for solidification of a liquid fermi liquid.
\newblock {\em arXiv preprint arXiv:2504.06508}, 2025.

\bibitem{leggett1970can}
Anthony~J Leggett.
\newblock Can a solid be" superfluid"?
\newblock {\em Physical Review Letters}, 25(22):1543, 1970.

\bibitem{andreev1969quantum}
AF~Andreev and IM~Lifshits.
\newblock Quantum theory of defects in crystals.
\newblock {\em Zhur Eksper Teoret Fiziki}, 56(6):2057--2068, 1969.

\bibitem{chester1970speculations}
GV~Chester.
\newblock Speculations on bose-einstein condensation and quantum crystals.
\newblock {\em Physical Review A}, 2(1):256, 1970.

\bibitem{kim2004probable}
Eunseong Kim and Moses Hung-Wai Chan.
\newblock Probable observation of a supersolid helium phase.
\newblock {\em Nature}, 427(6971):225--227, 2004.

\bibitem{kim2004observation}
Eunseong Kim and Moses~HW Chan.
\newblock Observation of superflow in solid helium.
\newblock {\em Science}, 305(5692):1941--1944, 2004.

\bibitem{anderson2005thermodynamics}
PW~Anderson, WF~Brinkman, and David~A Huse.
\newblock Thermodynamics of an incommensurate quantum crystal.
\newblock {\em Science}, 310(5751):1164--1166, 2005.

\bibitem{prokofev2005supersolid}
Nikolay Prokof'ev and Boris Svistunov.
\newblock Supersolid state of matter.
\newblock {\em Phys. Rev. Lett.}, 94:155302, Apr 2005.

\bibitem{boninsegni2012colloquium}
Massimo Boninsegni and Nikolay~V Prokof’ev.
\newblock Colloquium: Supersolids: What and where are they?
\newblock {\em Reviews of Modern Physics}, 84(2):759--776, 2012.

\bibitem{Tonghang_upcoming}
Tonghang Han, Jackson~P. Butler, Shenyong Ye, Zhenqi Hua, Surajit Dutta, Zach Hadjri, Zhenghan Wu, Jixiang Yang, Junseok Seo, Phatthanon Pattanakanvijit, Emily Aitken, Kenji Watanabe, Takashi Taniguchi, Peng Xiong, Eli Zeldov, Zhengguang Lu, Raymond Ashoori, and Long Ju.
\newblock Evidence of metallic wigner crystal in rhombohedral graphene.
\newblock 2026.

\bibitem{han2025signatures}
Tonghang Han, Zhengguang Lu, Zach Hadjri, Lihan Shi, Zhenghan Wu, Wei Xu, Yuxuan Yao, Armel~A. Cotten, Omid Sharifi~Sedeh, Henok Weldeyesus, Jixiang Yang, Junseok Seo, Shenyong Ye, Muyang Zhou, Haoyang Liu, Gang Shi, Zhenqi Hua, Kenji Watanabe, Takashi Taniguchi, Peng Xiong, Dominik~M. Zumbühl, Liang Fu, and Long Ju.
\newblock Signatures of chiral superconductivity in rhombohedral graphene.
\newblock {\em Nature}, 643(8072):654–661, May 2025.

\bibitem{nguyen2025hierarchy}
Ron~Q Nguyen, Hai-Tian Wu, Erin Morissette, Naiyuan~J Zhang, Peiyu Qin, Kenji Watanabe, Takashi Taniguchi, Aaron~W Hui, Dima~E Feldman, and JIA Li.
\newblock A hierarchy of superconductivity and topological charge density wave states in rhombohedral graphene.
\newblock {\em arXiv preprint arXiv:2507.22026}, 2025.

\bibitem{Note1}
We take $N$ and $N_{uc}$ to be continuous variables. Tiling problems at the boundary of the system will generate subextensive energy costs, and thus are of subleading concern.

\bibitem{Note2}
An alternative way to understand $k_0$ is as a Legendre transform of $E$ with respect to $N$ and $A_s$. The Legendre transform $N_{uc}k(\mu ,\protect \tilde {p},N_{uc})$ is continuous, while $N_\protect \text {uc}k(N,A_s,N_{uc})$ used in the text may be discontinuous.

\bibitem{Note3}
We define $k^\pm $ such that $k^+$ corresponds to when the system is slightly electron doped, and $k^-$ corresponds to when the system is slightly hole doped.

\bibitem{Note4}
The charge gap can be nonzero because we hold $N_{uc}$ fixed. If $N_{uc}$ is allowed to vary with $N$, the crystal can breathe to accommodate additional charges at no cost.

\bibitem{Note5}
We note that $E$ and $\partial E/\partial A_s$ are both continuous functions of $N,A_s,N_{uc}$. To show the second part of the statement, use Eq.~\protect \eqref {eq:eps_n_nuc} and $E=A_s\epsilon $. Since $N,N_{uc}$ are held fixed when taking the partial derivative, the derivative must be continuous at $N = N_{uc}$.

\bibitem{Zhihuan_upcoming}
Jiechao Feng, Zhaoyu Han, Michael~P. Zaletel, and Zhihuan Dong.
\newblock To appear.

\bibitem{auerbach2025isospin}
Nadav Auerbach, Surajit Dutta, Matan Uzan, Yaar Vituri, Yaozhang Zhou, Alexander~Y Meltzer, Sameer Grover, Tobias Holder, Peleg Emanuel, Martin~E Huber, et~al.
\newblock Isospin magnetic texture and intervalley exchange interaction in rhombohedral tetralayer graphene.
\newblock {\em Nature Physics}, 21(11):1765--1772, 2025.

\bibitem{footnote2}
Our interpretation relies on the assumption that $\delta n^*$ of the MWC does not significantly deviate from its zero-field value and the quantization field does not significantly change due to the Berry curvature hosted by the occupied states.

\bibitem{QTM}
A.~Inbar, J.~Birkbeck, J.~Xiao, T.~Taniguchi, K.~Watanabe, B.~Yan, Y.~Oreg, Ady Stern, E.~Berg, and S.~Ilani.
\newblock The quantum twisting microscope.
\newblock {\em Nature}, 614(7949):682–687, February 2023.

\bibitem{Note6}
We indeed confirm that this is the case for certain parameters in the phase diagram.

\end{thebibliography}

\onecolumngrid
\appendix
\newpage

\section{Table of variables related to thermodynamics}
Due to the large number of variables we employ for the thermodynamics, here we list a table of variables for the convenience of readers.

\begin{table}[htbp]
 \renewcommand{\arraystretch}{1.4}
 \centering
\begin{tabular}{ l l } 
 \toprule
  Symbol & Physical quantity  \\ \hline \\[-1.2em]
  $N$ & Number of electrons \\ 
  $A_s$ & Area of system \\
  $n=N/A_s$ & Electron density \\
  $A_{uc}$ & Area of crystalline unit cell \\
  $n_{uc}=A_{uc}^{-1}$ & Electron density commensurate with filling $1$ electron per unit cell area $A_{uc}$\\
  $\nu=nA_{uc}$ & Filling of electrons per crystalline unit cell \\
  $E$ & Energy  \\ 
  $N_{uc}=A_s/A_{uc} = N/\nu$ & Number of crystalline unit cells in system\\
  $k(N,A_s,N_{uc})$ & \tcell{Packing bias. Related to $E$ as $k=(E-N\partial_NE-A_s\partial_{A_s}E)/N_\text{uc}$}\\
  $\epsilon = E/A_s$ & Energy density per area \\
  $\Delta$ &  \tcell{The charge gap of a commensurate crystal; also cusp of $\epsilon$ on the commensurate line in both $n$ and $n_{uc}$ directions}\\
  $\delta n = n-m n_{uc}$ & \tcell{Deviation of density from the commensurate line with $m \in \mathbb{Z}$.}\\
 \bottomrule
\end{tabular}
\caption{Variables used in the thermodynamics of crystal commensurability.}
\label{tb:variables}
\end{table}

\section{Energy density and stability of commensurate crystals}
\label{app:CC}
We will now provide an equivalent derivation of Eq.~\eqref{eq:stability}. We will rewrite the energy functional $E(N,A_s, N_{uc})$ as
\begin{equation}
    E(N,A_s, N_{uc}) = A_s\epsilon\left(\frac{N}{A_s}, \frac{N_{uc}}{A_s}\right) = A_s\epsilon(n,n_{uc}).
\end{equation}
The functional $\epsilon(n,n_{uc})$ thus represents the ground state energy density per area of a crystalline state with density $n$ constrained to have crystal unit cell area $A_{uc}=n_{uc}^{-1}$. We focus on expanding around the commensuration line $n = m n_{uc},\nu \in\mathbb{Z}$, and we assume $\epsilon$ and charge gap $\Delta$ vary smoothly along the commensuration line.

As discussed in the main text, the commensurate crystal can develop a nonzero charge gap and exhibit a cusp singularity in $\epsilon$:~\cite{tesanovic_hall_1989}

\begin{equation}
\epsilon(n,n_{0})=\epsilon_0 + \frac{\Delta}{2} \left|n- m n_0\right| +\mu_0 (n- m n_0)+\dots.
\end{equation}
$\mu_0$ is the (smooth part of) chemical potential, and the charge gap is defined as
\begin{equation}
\label{eq:Delta_n}
    \Delta(n_0) = \frac{\partial \epsilon}{\partial n}(mn_0^{+},n_0) - \frac{\partial \epsilon}{\partial n}(mn_0^{-},n_0) = \mu^+ - \mu^-.
\end{equation}

When one allows $n_{uc}$ to freely vary, the energy density takes the form
\begin{equation}
\label{eq:energy_expansion}
    \epsilon(n,n_{uc})\approx \epsilon_0 + \frac{\Delta}{2} |n-m n_{uc}|+\mu_0(n-m n_0) + k_0(n_{uc}-n_0) 
\end{equation}
when both $n,n_{uc}$ are close to $n_0$. To see this, consider the following Clausius-Clapeyron-type argument: we write $\epsilon(mn_0 + m\delta, n_0 + \delta) $ in two different ways:
\begin{equation}
    \epsilon(mn_0 + m\delta, n_0 + \delta) = m\delta \partial_n \epsilon(mn_0^+, n_0) + \delta \partial_{n_{uc}}\epsilon(mn_0, n_0^-)
    = m\delta \partial_n \epsilon(mn_0^-, n_0) + \delta \partial_{n_{uc}} \epsilon(mn_0, n_0^+).
\end{equation}
This implies
\begin{equation}
     (\partial_{n_{uc}} \epsilon(mn_0, n_0^+) - \partial_{n_{uc}} \epsilon(mn_0, n_0^-)) = m\Delta,
\end{equation}
which is satisfied by the cusp singularity we introduced above.

The stability criterion for $n_{uc}$ is thus
\begin{equation}
\begin{aligned}
    \frac{\partial \epsilon}{\partial n_{uc}}(mn_0,n_0^+)&=-\frac{m\Delta  }{2}+k_0 \leq 0;\\
    \frac{\partial \epsilon}{\partial n_{uc}}(mn_0,n_0^-)&=\frac{m\Delta }{2}+k_0 \geq 0,
\end{aligned}
\end{equation}
which is exactly Eq.~\eqref{eq:stability}.
If either criterion is violated, the commensurate solution becomes energetically unstable, spontaneously changing the area of the unit cell to allow a metallic Wigner crystal.

We will now relate $k_0$ to thermodynamic properties of the commensurate crystal. Let $\tilde\epsilon = \epsilon - \frac{\Delta}{2}|n - m n_{uc}|$ denote the regular part of the energy density. We now compute the variation in energy density along the commensurability line:
\begin{equation}
    \frac{d}{dn} \tilde{\epsilon}(m n_0, n_0) \Bigg|_{n=\nu n_{uc}} = \partial_n \tilde{\epsilon}+m^{-1} \partial_{n_{uc}} \tilde{\epsilon} = \mu_0 + m^{-1} k_0.
\end{equation}
Since $n/n_{uc}=N/N_{uc}=m$, a change in $n$ on the commensurate line could equivalently be expressed as a change in the area of the system $A_s$ itself, while holding $N$ and $N_{uc}$ to be constant:

\begin{equation}
   \begin{aligned}
       \frac{d}{dn} \tilde{\epsilon}(m n_0, n_0) \Bigg|_{n=m n_{uc}}& = \frac{d A_s}{dn}\frac{\partial }{\partial A_s}\left(\frac{E(N,A_s,N_{uc})}{A_s}\right)
       = \frac{A_s}{N}\left(\frac{E}{A_s}-\frac{\partial E}{\partial A_s}\right)
       = \frac{\epsilon+\tilde{p}}{n}.
   \end{aligned} 
\end{equation}
Therefore, combining the two equations above, we obtain
\begin{equation}
    k_0=\frac{m (\epsilon-\mu_0 n+\tilde{p})}{n}=\frac{E-\mu_0 N+\tilde{p}A_s}{N_{uc}} 
\label{eq:k0}
\end{equation}

We finally comment on the possibility of continuous phase transitions from a commensurate WC to an incommensurate one. This could happen when, for example, $k_0$ is continuously tuned as it exceeds $\Delta/2$, where one develops an offset $n_{uc}-n_0$ proportional to $-\sgn(k_0)(|k_0|-\nu\Delta/2)$. This is akin in nature to a Lifshitz transition, where a chemical potential is continuously tuned through a band edge.

\section{Dependence of the phase diagram on interaction strength}
\label{app:eps_r_dependence}

\begin{figure}[htbp]
    \centering
    \includegraphics[width=\linewidth]{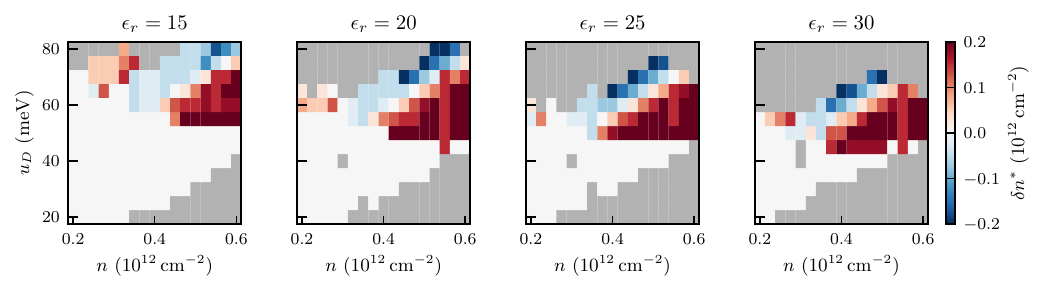}
    \caption{ Phase diagram evolution as a function of dielectric constant $\epsilon_r=15-30$.}
    \label{fig:delta_n_star_vs_eps_r}
\end{figure}
In Fig.~\ref{fig:delta_n_star_vs_eps_r} we plot the evolution of the phase diagram as we tune the dielectric constant $\epsilon_r$ from $15$ to $30$. We identify the following prominent features. Due to a decrease in interaction strength $U\propto \epsilon_r^{-1}$, the non-crystalline metal (NCM) region shown in gray expands in the phase diagram. Due to the same reason, the phase boundary between the insulating WC and the MWC phase moves down in $u_D$, as an increase in $u_D$ enhances $E_{\gamma}-E_{\kappa^+}$ due to the Mexican hat dispersion. However, despite the precise phase boundaries shifting, we emphasize that the broad trend remains the same. The MWC phase appears at a larger $u_D$ compared to the insulating WC phase. Furthermore, the diagonal stripe in $(n,u_D)$ plane where holes are favored rather than electrons remains robust. The results demonstrate that the MWC phase is not a numerically fine-tuned phase.

\section{Microscopic Model and Numerical Details}
\label{app:RMG}

We model rhombohedral graphene using the standard tight-binding model, following the conventions from~\cite{AHC1} and parameters from~\cite{auerbach2025isospin}. We reiterate the key details here to be self-contained.

\subsection{Single Particle Model}
We consider $M$-layer rhombohedral graphene with lattice $\v{R}_1 = (a,0), \v{R}_2 = (1/2,\sqrt{3}/2)a$ with atoms on sublattice $\sigma \in \{A,B\}$ and layer $\ell$ at $\v{r}_{\sigma,\ell} = (0, (\ell - \delta_{\sigma,A})/\sqrt{3},\ell c)$ where $a\approx \SI{2.46}{\r{A}}$ and $c$ is the interlayer spacing. We define nearest neighbor vectors $\v{\delta}_n = \mathsf{R}_{2\pi/3}^n (0,a/\sqrt{3})$ where $\mathsf{R}_{\theta}$ is the counterclockwise rotation matrix by $\theta$.

We work in a basis $\ket{\v{k}, \sigma,\ell} = \hat{c}^\dagger_{\v{k},\sigma,\ell}\ket{0}$. In terms of this, the single-particle Hamiltonian is $h_{\mathsf{RMG}}^{(M)}~=~\sum_{\v{k} \in \bz} \sum_{\ell,j} \hat{c}_{\v{k} \sigma j}^\dagger \left[ h_{RMG}^{(N_L)}(\v{k}) \right]_{\sigma j, \sigma'j'} \hat{c}_{\v{k} \sigma' j'} + \hat{U}_{\mathrm{disp}}$ where
represented by a block matrix in layer space,
\begin{equation}
	h_{\mathsf{RMG}}^{(M)}(\v{k}) =
	\begin{bmatrix} 
		h^{(0)} & h^{(1)} & h^{(2)}\\[0.5em]
		h^{(1)\dagger} & h^{(0)} & h^{(1)} & h^{(2)}\\[0.5em]
		h^{(2)\dagger} & h^{(1)\dagger} & h^{(0)} & \ddots & \ddots\\[0.5em]
				& h^{(2)\dagger} & \ddots & \ddots & h^{(1)} & h^{(2)}\\[0.5em]
				& & \ddots & h^{(1)\dagger} & h^{(0)} & h^{(1)}\\[0.5em]
				& & & h^{(2)\dagger} & h^{(1)\dagger} & h^{(0)}\\
	\end{bmatrix}
	\label{eq:rhombohedral_graphene}
\end{equation}
which is a matrix in layer space with entries in sublattice space
\begin{align}
    h^{(0)}_{\ell} (\v{k}) &= \begin{pmatrix} 0  & -t_0 f_{\v{k}}\\ -t_0 \overline{f}_{\v{k}} & 0 \end{pmatrix}, 
    &h^{(1)}(\v{k}) &= \begin{pmatrix} t_4 f_{\v{k}} & t_3 \overline{f_{\v{k}}} \\ t_1 & t_4 f_{\v{k}} \end{pmatrix},
    &h^{(2)}(\v{k}) &= \begin{pmatrix} 0 & \frac{t_2}{2}\\ 0 & 0 \end{pmatrix},
    &f_{\v{k}} &= \sum_{n=0}^2 e^{i \v{k}\cdot \v{\delta}_n}.
\end{align}

The on-site potential $\hat{U}_{\mathrm{disp}}$ term is a function of $u_D$, the potential difference between adjacent layers, which we approximate as proportional to the external displacement field. It is a diagonal operator of the form
\begin{align}
	\hat{U}_{\ell}(u_D) &= 
	 \sum_{\sigma}
	(u_1 + \Delta_2 + \delta_{\sigma,B} \delta) \hat{n}_{\v{k},\sigma,1} 
	+ \left(\sum_{\ell=2}^{M-1}
	(u_\ell - \Delta_2 + \delta) \hat{n}_{\v{k},\sigma,\ell}\right)
	+(u_M + \Delta_2 + \delta_{\sigma,A} \delta) \hat{n}_{\v{k},\sigma,M}
\end{align}
where the on-layer potentials are $u_\ell = u_D\left(\ell-1 - (M-1)/2\right)$. In the case of R4G, $M=4$, $u_{\ell} = u_D(-\frac{3}{2},-\frac{1}{2},\frac{1}{2},\frac{3}{2})_{\ell}$, so that the potentials are symmetric around the middle of the sample.

We take parameters following~\cite{auerbach2025isospin}.
\begin{center}
\begin{tabular}{ccccccc}
	\toprule
	$t_0$ (meV) & 
	$t_1$ (meV) & 
	$t_2$ (meV) & 
	$t_3$ (meV) & 
	$t_4$ (meV) & 
	$\delta$ (meV)	&
	$\Delta_2$ (meV) \\ 
	\midrule
	3100 & 380 & -15 & 290 & -141 & 10.5 & 2\\
	\bottomrule
\end{tabular}
\end{center}

Within the $K$ valley, the model has the following symmetries: three-fold rotation $C_3$;  mirror-x followed by time reversal $\mathcal{T} M_x = (k_x \to k_x, k_y \to -k_y) I_{\text{orb}} \mathsf{K}$ where $\mathsf{K}$ is complex conjugation and $I_{\text{orb}}$ is the identity matrix in orbital (sublattice,layer) space; and continuous translation symmetry (in the regime of one electron per thousand graphene unit cells or less that we consider).

As mentioned in the main text, we use a crystalline lattice generated by $\v{L}_1 = L(1,0)$ and $\v{L}_2 = L (-1/2,\sqrt{3}/2)$ and define the reciprocal lattice by $\v{g}_j \cdot \v{L}_i = 2\pi \delta_{ij}$. This choice preserves $C_3$ and $\mathcal{T} M_x$ symmetries. We note there is another symmetric choice of crystalline lattice. We have checked at isolated points that using the other crystalline lattice does not change our results appreciably. We define the following high-symmetry points of the crystalline lattice: 
\begin{equation}  
\begin{aligned}
	\v{\gamma} &= \v{K}_{\mathrm{Gr}},
	&\v{\kappa}^+ &= \v{\gamma} -\frac{1}{3}\v{g}_1 + \frac{1}{3}\v{g}_2,
	&\v{\kappa}^- &= \v{\gamma} -\frac{2}{3}\v{g}_1 - \frac{1}{3}\v{g}_2, 
	&\v{\mu} &= \v{\gamma} - \frac{1}{2}\v{g}_1,
    &\v{\mu}' &= \v{\gamma} + \frac{1}{2}\v{g}_1.
\end{aligned}
\end{equation}

\subsection{Many-Body Model}

As noted in the main text, we use
\begin{equation}
    \label{eq:Hamiltonian_app}
    \hat{H} = \hat{h}_{\mathsf{RMG}} + \frac{1}{2 A_s} \sum_{\v{q}} V_{\v{q}} :\hat{\rho}_{\v{q}} \hat{\rho}_{-\v{q}}:, \quad
    V_{q} = \frac{2\pi \tanh q d}{\epsilon_r \epsilon_0 q},
\end{equation}
where $\hat{\rho}_{\v{q}}$ is the density operator at momentum $\v{q}$, and normal ordering is relative to the charge-neutral gap of RMG, which is a natural ``subtraction scheme" in the limit of a large gap at neutrality. 

We solve Eq.~\eqref{eq:Hamiltonian_app} using SCHF following, e.g.,~\cite{AHC1}. We project the Hamiltonian into the lowest $N_b = 13$ conduction bands, which is sufficient for good convergence of most metallic states under consideration. To reliably find the global ground state, we take at least ten different ``seed" ground states for each Hamiltonian, eventually selecting the one with the lowest energy. However, states with very small Fermi pockets can have small basins of attraction, making them hard to find even with many seeds. We use ODA to achieve convergence, as DIIS is often unstable for metallic states. In practice we achieve convergence to the $\dn{[h_{HF}[P], P]}_2 < \SI{e-3}{meV/electron}$ level, and often much better.

\section{Properties of Metallic Wigner Crystal States}

\begin{figure*}[htbp]
    \centering
    \includegraphics[width=\linewidth]{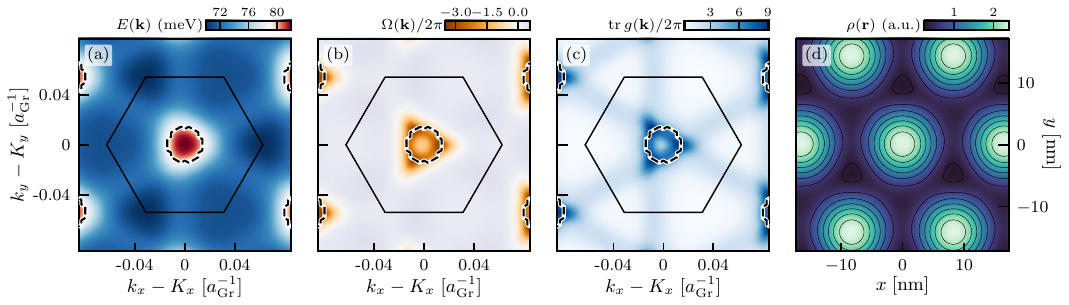}
    \caption{
    Properties of the metallic Wigner crystal state.
    (a) Energy of the lowest SCHF band. The Fermi surface of holes is marked with a dashed black line.
    (b) Berry curvature of the lowest SCHF band, normalized so the Chern number is $C = A_{\mathrm{bz}}^{-1}\int_{BZ} (\Omega(k)/2\pi)$.
    (c) Trace of the quantum metric with the same normalization.
    (d) Real-space charge density of the metallic Wigner crystal. 
    Parameters:
    $(n/n_{12},u_D,\epsilon_r,\delta n/n_{12}) = (0.4, 60\text{meV},15,-0.025)$
    where $n_{12} = \SI{e12}{\per\centi\meter\squared}$.
    }
    \label{fig:MWC_detailed_properties}
\end{figure*}

Detailed properties of the MWC shown in Fig.~\ref{fig:MWC_properties} are demonstrated in Fig.~\ref{fig:MWC_detailed_properties}. Fig.~\ref{fig:MWC_detailed_properties}(a) shows the band structure of the lowest energy band, with a small Fermi pocket of holes at $\gamma$. The state is $C_3$ and $\mathcal{T} M_x$ symmetric.
Fig.~\ref{fig:MWC_detailed_properties}(b) shows the Berry curvature $\Omega(\v{k})$ of the lowest band. Away from $\gamma$, the Berry curvature is positive, reflecting the ``intrinsic" Berry curvature in the $K$ valley of RMG. Near $\gamma$, avoided crossings with the higher bands, shown in Fig.~\ref{fig:MWC_properties}(a,b), create a triangle of ``extrinsic" Berry curvature. Crucially, the sign of the extrinsic Berry curvature is \textit{opposite} to that of the parent band --- and fairly large. Finally, Fig.~\ref{fig:MWC_detailed_properties}(d) shows the charge density profile for the occupied states. Superficially, it is akin to the insulating Wigner crystal, with a triangular lattice of localized charge density peaks.

\begin{figure}
    \centering
    \includegraphics[width=\linewidth]{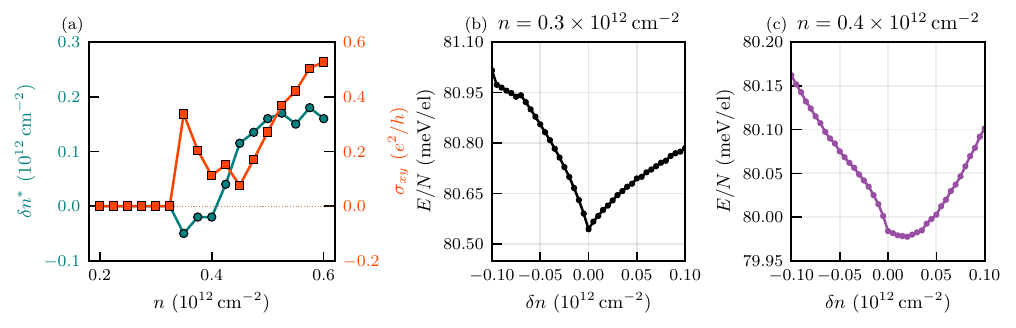}
    \caption{
    (a) Optimal incommensuration $\delta n^*$ and Hall conductivity as a function of density at $u_D = \SI{60}{meV}$. Parameters match Fig.~\ref{fig:Fermi_surfaces}. (b, c) $E(\delta n)/N$ as a function of $\delta n$ at $u_D=\SI{60}{meV}$, $n=(0.3,0.4)\times\SI{e12}{\per\centi\meter\squared}$ respectively. A cusp is clearly visible at $\delta n=0$ in (b) and a tilted cusp at $\delta n=0$ in (c).} 
    \label{fig:conductivity_vs_density}
\end{figure}

Although the lowest band is topologically trivial, the significant Berry curvature can give rise to a large Hall conductivity. The state in Fig.~\ref{fig:MWC_detailed_properties} exhibits Hall conductance $|\sigma_{xy}|\approx 0.1 e^2/h$. Fig.~\ref{fig:conductivity_vs_density}(a) examines the Hall conductivity as a function of density, finding it is often a significant fraction of $e^2/h$ in the electron-doped regime of the MWC. Two representative points $n=(0.3,0.4)\times\SI{e12}{\per\centi\meter\squared}$ are chosen and their respective $E(\delta n)/N$ is shown in Fig.~\ref{fig:conductivity_vs_density}(b,c), demonstrating clear cusps at $\delta n=0$. The cusp is robust enough to provide stability for Fig.~\ref{fig:conductivity_vs_density}(b), but not enough against the large packing bias in Fig.~\ref{fig:conductivity_vs_density}(c).

\section{TDHF}
We further carry out time-dependent Hartree-Fock (TDHF) on some of the ground states, following~\cite{AHC4}. We find it is crucial to start from a well-converged ground state with $C_3$ and $\mathcal{T} M_x$ symmetry to avoid spurious low energy modes. For carrying out TDHF on such large system sizes, it is crucial to apply the TDHF operator as a linear function and use Krylov techniques. Doing this, we converge the ${\sim} 20$ eigenvalues to nearly machine precision with the smallest real part at each relative momentum $\v{q}$.

\begin{figure}[htbp]
    \centering
    \includegraphics[width=0.5\linewidth]{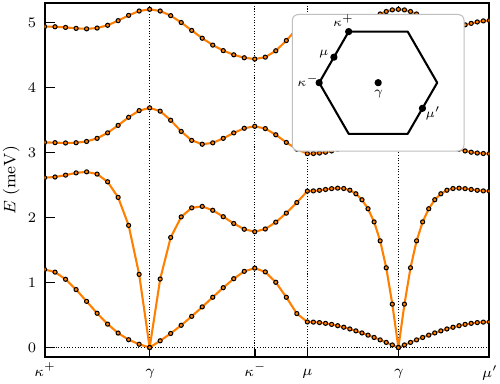}
    \caption{
    Time-dependent Hartree-Fock spectrum in the insulating Wigner crystal state. The gapless bands at $\gamma$ are the longitudinal and transverse phonon modes of the Wigner crystal.
    Parameters:
    $(n,u_D,\epsilon_r,\delta n) = (\SI{0.4e12}{\per\centi\meter\squared}, 50\si{meV},15,0)$.
    }
    \label{fig:TDHF_WC}
\end{figure}

Fig.~\ref{fig:TDHF_WC} shows a representative TDHF spectrum for a Wigner crystal. As expected, there are two gapless modes: the transverse and longitudinal phonons. The speed of sound $v_s$ is highly directional, and significantly more pronounced than the MWC state shown in the main text in Fig.~\ref{fig:TDHF_MWC}. Along $\gamma \to \kappa^+$ we estimate $v_s \approx \SI{3.1}{km/s}$, while from $\gamma \to \kappa^-$ we find $v_s \approx \SI{6.4}{km/s}$, more than double the previous value. This is due to a large kineoelastic term allowed in the absence of inversion and time-reversal~\cite{AHC4}.

\section{Detailed analysis on the stability of commensurate Crystals}
\label{app:analysis_stability}

\begin{figure}[htbp]
    \centering
    \includegraphics[width=0.8\linewidth]{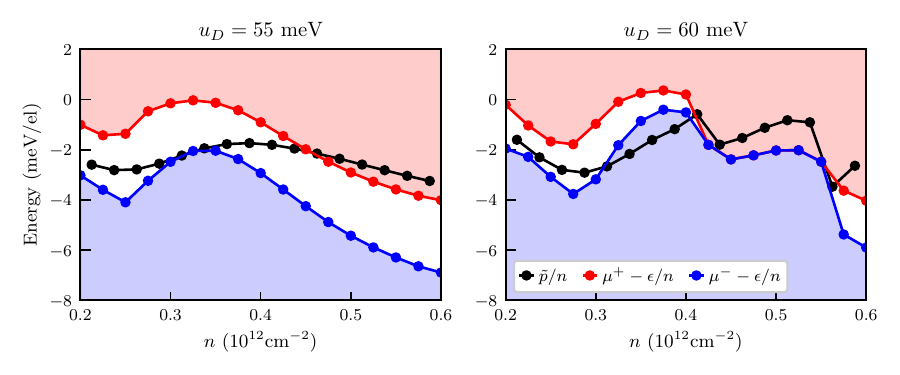}
    \caption{Dependence of $\tilde{p}/n$ (black) and $\mu^\pm-\epsilon/n$ (red, blue) as a function of $n$ for $u_D=55,60\si{meV}$. When $\tilde{p}/n$ is above $\mu^+-\epsilon/n$, the system is unstable to an expansion of unit cells, thus introducing electron-like carriers. When $\tilde{p}/n$ is below $\mu^--\epsilon/n$, the system is unstable to a contraction of unit cells, thus introducing hole-like carriers. The commensurate crystal is stable when $\tilde p/n$ is between $\mu^\pm-\epsilon/n$.}
    \label{fig:components_K}
\end{figure}

As shown in Eq.~\eqref{eq:stability} and in App.~\ref{app:CC}, the stability of the commensurate crystal is governed by the difference in magnitude between $\Delta/2$ and $k_0$. Using Eq.~\eqref{eq:k0}, one can recast the stability condition as
\begin{equation}
   \mu^--\frac{\epsilon}{n}\leq \frac{\tilde{p}}{n} \leq \mu^+-\frac{\epsilon}{n}
\end{equation}
in which $\mu^{\pm}=\mu_0\pm\Delta/2$. $\mu^\pm - \epsilon/n$ can be estimated from a single Hartree-Fock calculation using Koopmans' theorem, and $\tilde{p}$ can be computed either by the G\"uttinger-Hellmann-Feynman theorem, or by taking a numerical derivative.
In Fig.~\ref{fig:components_K} we plot $\tilde{p}/n$ (black) and $\mu^\pm-\epsilon/n$ (red, blue) as a function of $n$ for $u_D=55,60\si{meV}$. When $\tilde{p}/n$ crosses into the red region above $\mu^+-\epsilon/n$, the system expands its unit cell, introducing electron-like carriers, and vice versa. This criterion accurately predicts the sign of the carriers in the metallic Wigner crystals in the phase diagram Fig.~\ref{fig:phase_diagram}. Curiously, this analysis reveals that the commensurate Wigner crystal at $\sim \SI{0.35e12}{\per\centi\meter\squared}$ is close to being unstable to self-doping hole-like carriers.

At $u_D = 55\si{meV}$, the pressure $\tilde p$ stays relatively constant, while $\mu - \epsilon/n$ changes rapidly. At $u_D = 60\si{meV}$, both $\tilde p$ and $\mu-\epsilon/n$ change rapidly, but, comparatively, $\mu^\pm-\epsilon/n$ exhibits larger fluctuations than $\tilde{p}/n$.

\end{document}